\documentclass[aps ,prl,twocolumn,superscriptaddress,nofootinbib,10pt,preprintnumbers,longbibliography
]{revtex4-1}

\usepackage[colorlinks=true,linkcolor=black,citecolor=blue,urlcolor=blue, pdfborder={0 0 0}]{hyperref}
\usepackage{mathtools}
\usepackage[utf8]{inputenc}
\usepackage{color}
\usepackage[table,xcdraw,dvipsnames]{xcolor}
\usepackage{multirow}
\usepackage[normalem]{ulem}    
\usepackage{mathtools}
  \usepackage{amsmath}
  \usepackage{amssymb}
\def\eq#1{{Eq.~(\ref{#1})}}
\def\eqs#1#2{{Eqs.~(\ref{#1}) and (\ref{#2})}}

 \def\G#1{\mathcal{G}_{#1}}

\newcommand{\Sec}[1]{ \medskip \noindent {\bfseries  #1}.}

\newcommand{\Subsec}[1]{ \medskip \noindent 
$\mathbf{\bullet}$ {\itshape #1}.}

\definecolor{ultramarine}{RGB}{0,32,96}

\renewcommand{\bar}{\overline}

\newcommand{\mX}{\mathcal{X}}

\newcommand{\UPQ}{U(1)_{\mathrm{PQ}}}
\newcommand{\mP}{m_{\mathrm{P}}}
\newcommand{\mQ}{\mathcal{Q}}
      \def\vev#1{\langle #1\rangle}

\def\Tr{\mbox{Tr}\,}
\def\Min{\mbox{Mnr}\,}
\def\diag{\mbox{diag}\,}
\def\det{\mbox{det}\,}
\def\arg{\mbox{arg}\,}

\newcommand{\beq}{\begin{equation}}
\newcommand{\eeq}{\end{equation}}
\newcommand{\bea}{\begin{eqnarray}}
\newcommand{\eea}{\end{eqnarray}}

\newcommand{\eqn}[1]{Eq.~(\ref{#1})}

\def \lsim{\mathrel{\vcenter
     {\hbox{$<$}\nointerlineskip\hbox{$\sim$}}}}
\def \gsim{\mathrel{\vcenter
     {\hbox{$>$}\nointerlineskip\hbox{$\sim$}}}}
     
\def\PQv{{\begin{picture}(13,0)(0,0)\put(0,0){\rm
PQ}\put(0,0){\line(2,1){14.5}}\end{picture}}~}

\begin{document}

\title{Exact accidental $\mathbf{U(1)}$ symmetries for the axion}

\newcommand{\affINFN}{{\small \it INFN, Laboratori Nazionali di Frascati, C.P.~13, 100044 Frascati, Italy}} 

\author{Luc Darm\'e}
\email{luc.darme@lnf.infn.it}
%
\author{Enrico Nardi}
\email{enrico.nardi@lnf.infn.it}
\affiliation{\affINFN}

\begin{abstract}
We study a class of gauge groups that can automatically yield 
a  perturbatively exact Peccei-Quinn symmetry, and we outline a model in which the 
axion quality problem is solved at all operator dimensions.  Gauge groups 
belonging to this class can also enforce and protect accidental  symmetries 
of the clockwork  type, and we present a toy model  where an  `invisible' axion   
arises from a single breaking of  the gauge and global symmetries.  
\end{abstract}

 \maketitle

\parskip 2pt

\label{sec:intro}
\Sec{Introduction} 
The non-trivial structure of the vacuum of Yang-Mills theories~\cite{Belavin:1975fg} implies that CP violation is a 
built-in feature in QCD~\cite{Callan:1976je,Jackiw:1976pf}.  Strong CP violation  is parametrized in 
terms of an angular variable $\theta \in [0,2\pi]$  whose  value is not determined by the theory,  but  
 is experimentally  bounded to lie surprisingly  close  to zero $|\theta| \lsim 10^{-10}$.  It is hard to believe that 
this could occur simply as whim of nature, especially because any value $\theta \lesssim 10^{-1}$  would leave our 
Universe basically unaffected~\cite{Ubaldi:2008nf,Dine:2018glh,Lee:2020tmi}, precluding  an anthropic explanation.  
A convincing rationale for $\theta \approx 0$ is provided by the Peccei-Quinn (PQ) 
mechanism~\cite{Peccei:1977hh,Peccei:1977ur}, which postulates the existence 
of a global Abelian symmetry,  endowed  with a mixed $\UPQ$-$SU(3)_{C}$ anomaly and broken 
spontaneously. This  unavoidably implies a  quasi-massless spin zero boson,  
\emph{the axion}~\cite{Weinberg:1977ma,Wilczek:1977pj}, whose central role is to relax  
dynamically $\theta$ to $0$. 
Remarkably, the axion also provides a novel solution to the apparently unrelated puzzle of the origin of 
dark matter~\cite{Abbott:1982af,Dine:1982ah,Preskill:1982cy},  as well as a plethora of other implications 
for astrophysics and cosmology (for a recent review see~\cite{DiLuzio:2020wdo}). 
However, it also raises various new issues. Among the deepest new questions stands the very 
origin of the axion or, more precisely,  which  is {\it `the origin of the PQ symmetry'}\,? 
There are in fact good reasons to believe that global symmetries cannot be fundamental, and 
this is especially true for a symmetry that, being anomalous, does not survive at the quantum level. 
A satisfactory explanation  would arise if, in some suitable extension of the Standard Model (SM),   
the  PQ symmetry occurs accidentally, in the sense that all renormalizable Lagrangian terms respecting 
first principles (Lorentz and local gauge invariance)  preserve automatically also a global $U(1)$ with 
the required properties. 
A second problem emerges because to comply with the  bound  $|\theta| < 10^{-10}$, 
 $\UPQ$ must be respected by all effective  operators acquiring a vacuum expectation value (VEV)  up to 
dimension $D\gsim11$. This is at odd with the well founded belief that all global symmetries are eventually 
violated by operators of all types and dimensions  induced by  quantum  
gravity~\cite{Dine:1986bg,Barr:1992qq,Kamionkowski:1992mf,Holman:1992us,Ghigna:1992iv,
Randall:1992ut,Abbott:1989jw,Coleman:1989zu,Kallosh:1995hi,Alonso:2017avz,Alvey:2020nyh}.
This is known as   {\it `the PQ symmetry quality problem'}.
A third issue is related with {\it `the axion scale'}. 
The axion is a periodic field that, to comply with phenomenological constraints, 
must take values over a compact space of rather large radius $v_a\sim 10^{10\pm 2}\,$GeV.  
In benchmark  models this is generally engineered by identifying $v_a$ with the 
PQ spontaneous symmetry breaking (SSB) scale $v_{\rm PQ}$. 
 This, however, brings in the usual  problem of stabilising the electroweak scale 
 against $\mathcal{O}(v_{\rm PQ})$ corrections. 
Various strategies have been put forth to explain  the origin of the PQ symmetry and  
protect it up to  a suitable operator dimension $D$:  discrete gauge symmetries  
$\mathbb{Z}_D$~\cite{Krauss:1988zc,Dias:2002gg,Carpenter:2009zs,Harigaya:2013vja,Dias:2014osa,Harigaya:2015soa,Ringwald:2015dsf},
multiple scalars with values of  $U(1)$  gauge charges of order $D$~\cite{Barr:1992qq}, 
non-Abelian gauge symmetries, which generally  have degree not less than~$D$~\cite{DiLuzio:2017tjx,Ardu:2020qmo},  
often  assisted by  supersymmetry~\cite{Lillard:2017cwx,Lillard:2018fdt,Nakai:2021nyf} 
or by higher dimensional constructions~\cite{Hill:2002me,Hill:2002kq,Redi:2016esr,Yamada:2021uze}.
However,  an unsatisfactory aspect of all these solutions is that if the scale of PQ-breaking ($\PQv$)  effects 
lies below $\mP$,  if PQ  SSB 
occurs at  a scale $v_{\rm PQ} \gg 10^{10}\,$GeV, or if future experimental limits 
will  hint to  $\theta \ll 10^{-10}$,  the  value of $D$  will have to be accordingly increased.   
As regards  the axion scale problem, certain solutions have been attempted exploiting the 
so-called  clockwork mechanism~\cite{Choi:2015fiu,Kaplan:2015fuy,Giudice:2016yja,
Ahmed:2016viu,Craig:2017cda,Giudice:2017suc,Choi:2017ncj,Giudice:2017fmj,Teresi:2018eai}.  
Clockwork PQ symmetries  allow to boost selectively some axion 
couplings~\cite{Farina:2016tgd,DiLuzio:2017pfr,Darme:2020gyx,Choi:2020rgn}, and to 
exponentially enhance~\cite{Coy:2017yex} or suppress~\cite{Bonnefoy:2018ibr}  the 
ratio $v_a/v_{\rm PQ}$. 
Clearly,  also  these symmetries call for an explanation of their   
 {\it origin} and required {\it high quality}. 
 However,  devising ways to  generate and protect  clockwork symmetries
 employing   first principles  is  an even more challenging task. 

In this Letter we show that a so far uncharted type of    
{\it flavor}\,\footnote{In this work `flavor' refers 
to a replication of exotic quarks.} gauge symmetries of the 
form $\G{MN}$$\,=\,$ $SU(M)\!\times\! SU(N)$   with $M$$\,\neq\,$$N$,  
that we henceforth denote as  `rectangular'  symmetries,  
allow to solve at the root  the  PQ origin and quality problems, by enforcing automatically 
global $U(1)$  symmetries that are either  perturbatively exact at the Lagrangian level, or that 
become {\it exact on the vacuum}.\footnote{This term refers to  
global symmetries that are broken explicitly solely   by operators  whose  VEV vanishes.  
Vacua having  more global symmetries  than   the Lagrangian yield  additional massless scalars  
besides the usual Nambu-Goldstone-Bosons (NGB)~\cite{Georgi:1975tz}.}
We outline a simple example where axion protection is enforced by 
 $SU(4)\!\times\! SU(2)$.
Finally, we  speculate how rectangular symmetries might  prove useful to solve also the axion scale problem.  
To illustrate this we construct a toy model wherein a clockwork  PQ symmetry arises automatically, and a large  
axion scale $v_a$  results from a  
gauge/global symmetry spontaneously broken by~VEVs~$v\ll v_{a}$. \\ [-16pt]

\label{sec:scalar}
\Sec{Rectangular gauge groups and accidental $\mathbf{U(1)_{\rm\bf PQ}}$}
Consider a scalar multiplet $Y$ transforming in the bi-fundamental representation $(M,\bar N)$  of the
gauge group $\G{MN} \!= \!SU(M)\times SU(N)$ with  $M\!>\! N\! \geq\! 2$. Let~us denote a generic  component 
as $Y_{\alpha i}$  where Greek indices span  $SU(M)$ and Latin indices   $SU(N)$.   
Each group factor has a pair of  Kronecker and Levi-Civita  
invariant tensors  $(\delta_M,\epsilon_M)$,   $(\delta_N,\epsilon_N)$
which can  be used to  construct  
invariants  by contracting  the indices of field components.
The renormalizable Lagrangian always contains the two  invariants 
$T\equiv \Tr(Y^\dagger Y)$   
and  $T_4\equiv \Tr(Y^\dagger Y)^2$ constructed from $\delta_M$ 
and $\delta_N$. Being~Hermitian,~they~are~\mbox{manifestly} invariant under a global $U(1)_{\xi_Y}$  phase redefinition $Y\!\to\! e^{i\xi}Y$.
Let us denote the trace of the matrix of the minors of order $k$ of $Y^\dagger Y$ 
as  $C_k=\Tr[\Min(Y^\dagger Y,k)]$. 
We have $T=C_1$ and we   replace $T_4$ with    
$A=\frac{1}{2}(T^2-T_4)= C_2 $~\cite{Nardi:2011st}. 
The $C_k$'s up to  $C_N= \det[Y^\dagger Y]$  form a fundamental set of 
  $\G{MN} \times U(1)_{\xi_Y}$ invariants: 
it can be proven~\cite{Espinosa:2012uu} that 
any higher order invariant $T_{2k} = \Tr(Y^\dagger Y)^k$ can be expressed in terms of  
this set.\footnote{Considering the $M\times M$ matrix  $YY^\dagger$ 
one has $\Tr[\Min( YY^\dagger,k)] = C_k $ 
for $k\leq N$ and 0 for $k>N$
which yields the same result.}
The accidental 
$U(1)_{\xi_Y}$ can only be broken by non-Hermitian invariants,   that are monomials with an 
unequal number of $Y$ and $Y^\dagger$ components, which must then involve the $\epsilon$ tensors. 
However,  all invariants involving $\epsilon_M$ and a single scalar multiplet vanish symmetrically. 
Consider in fact the $SU(M)$ singlet 
\begin{align}
\label{eq:epsid}
    \epsilon^{\alpha_1\dots \alpha_M} Y_{\alpha_1, i_1} \dots Y_{\alpha_M, i_M}\equiv (\epsilon_{M} Y^{M})_{i_1\dots i_M}\,,
\end{align}
where the right hand side (r.h.s) defines a shorthand notation  
for the contraction of  $SU(M)$ indices with $\epsilon_M$.   Since $M>N$  at 
least two components have the same $SU(N)$ index, so that 
the string vanishes symmetrically.\footnote{Only  for `square' symmetries with  $M=N$ are $U(1)_{\xi_Y}$-breaking 
operators like $\epsilon_M\epsilon_N Y^M \propto \det Y$ allowed~\cite{Nardi:2011st,Fong:2013sba,DiLuzio:2017tjx}.}
Thus the Lagrangian for a scalar multiplet $Y$  transforming under a rectangular gauge 
 symmetry  automatically enjoys a global  $U(1)_{{\xi_Y}}$ which is perturbatively  \underline{\it exact}.

To promote $U(1)_{\xi_Y}$ to a PQ symmetry,  it must be  
endowed with  a QCD anomaly. This requires 
assigning $U(1)_{\xi_Y}$  charges to fermions that 
carry color and couple to  $Y$.
Let us introduce two sets of chiral exotic  quarks in the fundamental 
of $SU(3)_C$, singlets under the electroweak gauge group, and transforming under $\G{MN}$ as 
$\mQ_L\sim (M,1)$ and  $Q_R\sim (1,N)$
so that the  Yukawa  operator  $\bar \mQ_L Y Q_R $ is gauge invariant.
To prevent a color gauge anomaly we  add $P=M\!-\!N$ quarks 
$q_R$, and a new scalar multiplet $Z$  acquiring a VEV 
so that all the quarks can be massive.
 This step can be arranged in different ways, the two extreme possibilities are:
\begin{itemize}
    \item[(I)]  Add a set of  $\G{MN}$-singlets $q_{Ra}$ ($a=1,\dots,P$)  
    which couple to a scalar multiplet  $ Z \sim (M,1)$ via $P$ 
    Yukawa operators  $\sum_{a=1}^P \bar \mQ_L \,Z\, q_{Ra}$. 
    \item[(II)] Assign the $q_R$'s to  the fundamental representation  
    of a new gauge factor $SU(P)$, and  $Z$   
     to the bi-fundamental  $(M,\bar P)$ of $\G{MP}$, so that there is a single  Yukawa operator $\bar \mQ_L\, Z\, q_R$.
\end{itemize}
Note that for $M=N+1$ the two cases coincide, hence we restrict case (II) 
to $P\geq 2$. $G_{MN(P)}$ gauge anomalies 
can be canceled  by adding three copies of $M,N,(P)$-plets of colorless `leptons' 
of  chirality opposite to that of the quarks, which can acquire mass  from the VEVs of the 
same   multiplets  $Y$ and $Z$, e.g.  $\sum_{r=1}^{3}\bar{\mathcal{L}^{r}_{R}} Y \ell^{\,r}_L $ etc.

Scalar  terms   involving only  $Z$     also enjoy 
an exact accidental symmetry  $U(1)_{\xi_{Z}}$, i.e. $V(Z) = V(Z^\dagger Z)$.
However, by contracting the $SU(M)$ indices of   $Y$ and $Z$ it is  
possible to construct certain  mixed non-Hermitian  operators that    
break  $U(1)_{\xi_{Y}}\times U(1)_{\xi_{Z}}$ to a single $U(1)$, that is defined 
by  some specific condition between the $U(1)$ charges $\mX_{Y}$ and $\mX_{Z}$.  
As it will become clear below, depending if  $SU(M)$ index contraction  is 
performed with $\delta^\alpha_{\beta}$  or  with $\epsilon^{\alpha_1\dots \alpha_M}$ 
the two possibilities are
\begin{eqnarray}
\label{eq:U1a}
\delta_M\!: && 
\  U(1)_{\xi_Y}\times U(1)_{\xi_Z} \to U(1)_{\xi}, \quad \ \ 
\mX_{Y}-\mX_{Z}=0\\
\label{eq:U1b}
\epsilon_M\!: && 
\  U(1)_{\xi_Y}\times U(1)_{\xi_Z} \to U(1)_{\xi'}, \ 
N\mX_{Y}+ P\mX_{Z}=0.\ \ 
\end{eqnarray}
The charge relation in \eqn{eq:U1b} implies that  $U(1)_{\xi'}$
has no QCD anomaly. Hence the  symmetry 
preserved by the operators constructed with $\epsilon_M$ cannot be 
promoted to a PQ symmetry.  
To see this let us consider a chiral transformation with generic 
quark charges $\mX_{\mathcal{Q}_{L}},\mX_{Q_{R}}, \mX_{q_{R}}$.
The $U(1)$-QCD anomaly coefficient is    precisely 
\begin{equation}
\label{eq:2N}
|2\mathcal{N}| = M\mX_{\mathcal{Q}_{L}} - N\mX_{Q_{R}} 
 - P \mX_{q_{R}} = N\mX_{Y}+ P\mX_{Z}\,, 
\end{equation}
where the relation with the charges of the scalars follows 
from requiring $U(1)$ invariance of the Yukawa terms.

 \label{sec:PQbreaking}
\Subsec{$U(1)$-breaking operators} 
 \eqs{eq:U1a}{eq:U1b} show that  
 operators involving  $\delta_M$  break  $U(1)_{\xi'}$, while  $\epsilon_M$-type of operators break  $U(1)_{\xi}$, so that  
in the presence of both no $U(1)$ would survive.  
To see which  
operators can arise, let us start with case (I) where  the multiplets  have components $Y_{\alpha\, i},\,Z_{\alpha}$.  Let us define 
a  set of  $SU(N)$ vectors  
$(X_n)_i=(Z^\dagger (Y Y^\dagger)^{n-1} Y)_i$, $n=1,\dots, N$. 
The operator 
\begin{equation}
\label{eq:O1Xn}
\mathcal{O}_I(X_n) = \epsilon_N \,\Pi_{n=1}^{N}\, X_n   
\end{equation}
does not vanish symmetrically, is  non-renormalizable ($D=N(N+1)\geq 6$ for $N\geq2$) and preserves $U(1)_{\xi}$.
Since for  $M-N\geq 2$ all $\epsilon_M$  contractions must involve at least two $Z_\alpha$, they vanish symmetrically, and thus 
$U(1)_\xi$  survives as a perturbatively  exact accidental symmetry,  broken  only
by the  anomaly with coefficient $|2\mathcal{N}|= (N+P)\mX_{Y}$. 
For $M\!-\!N=1$ instead we can   write\nobreak   
\begin{equation}
\label{eq:O1YZ}
\mathcal{O}'_I(Y,Z) = (N!)^{-1}\,\epsilon^{\alpha_1\dots \alpha_{N}\alpha_M}
\left(\epsilon_N Y^N\right)_{\alpha_1\dots \alpha_N} Z_{\alpha_M} \,,
\end{equation}
that has dimension $D=M$ (and hence is renormalizable for $\G{32}$ and $\G{43}$).
Then, in this particular case  $U(1)_{\xi'}$ gets broken at $D=M\cdot N$
and no protected $U(1)$ survives.
%
%
 
\begin{table}[t]
 \renewcommand{\arraystretch}{1.3}
\begin{center}
\footnotesize
\begin{tabular}{lcc}
\hline
\hline
\qquad Case &  $D(\mathcal{O}')$ 
\ \ &\  \  $D(\mathcal{O})$ 
\ \ \ \ \  \  \\
\hline
(I)\  \  $M-N=1$ &\hspace{1cm}   $M$\hspace{1cm}  & $ N(N+1)$ \\
\phantom{(I)}\ \ $M-N>1$ & $-$ & $N(N+1)$  \\
 (II)\ $N=P$ & $M$ & $M$  \\
\phantom{(II)}\ $N>P$ & $M$  & $D(L)$  \\
\hline
\hline
\end{tabular}
\end{center}
\vspace{-0.4cm}
\caption{Dimension of the operators  
 $\mathcal{O}'$ and  $\mathcal{O}$ of lowest order 
 that break respectively  $U(1)_{\xi }$ and  $U(1)'_{\xi }$.
The expression for $D(L)$ is given in the text. 
}
\label{tab:Dim}
\vspace{-0.3cm}
\end{table}

In   case (II) the  multiplets  components  are
 $Y_{\alpha i},\,Z_{\alpha a}$ where  
 $a,b,\dots$ span $SU(P)$. 
 Let us take $N\!\geq\! P\geq\!2$ ($N\leq P$ amounts to interchange $Y\leftrightarrow Z$)
and let us consider the $SU(P)$ and $SU(N)$ singlets $(\epsilon_P Z^P)_{\alpha_1\dots \alpha_P}$ and  
 $(\epsilon_N Y^N)_{\beta_{1}\dots \beta_{N}}$.
Since  $M=P+N$ the  $SU(M)$ indices of their product 
can be exactly saturated with  $\epsilon_M$,  yielding the $\G{MNP}$   invariant operator 
 of dimension $D=M$ 
\begin{equation}
\label{eq:O2YZ}
    \mathcal{O}'_{I\!I}(Y,Z) = (P!\, N!)^{-1}\,
    \epsilon_M \,
        \left(\epsilon_P Z^P\right) \,
            \left(\epsilon_N  Y^N\right)\,, 
\end{equation}
 that preserves  $U(1)_{\xi'}$ (and  is renormalizable for $\G{422}$).
$\delta_M$-type of operators can be \mbox{constructed} 
starting  from  $(\epsilon_P Z^{\dagger P})^{\alpha_1\dots \alpha_P}$  and by 
contracting the  $SU(M)$ indices with $P$ components of $Y$. 
Defining $(X_1)^a_i = (Z^\dagger Y)^a_i$
this   yields $(\epsilon_P X_1^P)_{i_1,\dots i_P}$. 
The  $SU(N)$ indices can be contracted with $\epsilon_N$ only 
if $N=P=M/2$, that is when  $X_1$  is a  $N\times N$  square matrix. The $D=M$ operator 
\begin{equation}
\label{eq:O2X}
\mathcal{O}_{I\!I}(X_1) = (P!)^{-1}\,\epsilon_N \epsilon_P X_1^N  =  \det X_1 \,, 
\end{equation}
is also renormalizable only for $\G{422}$, and  is 
invariant under $U(1)_\xi$. 
For  $P< N \leq 2P $,  adding $N-P$ new objects $(X_2)^a_i= ( Z^\dagger Y Y^\dagger Y)^a_i$
allows for the contraction  
$[\epsilon_N (\epsilon_P X_1^P) {X_2}^{N-P}]^{a_1\dots a_{N-P}}$.\footnote{Different field combinations $X_2\neq X_1$ are  needed because 
otherwise $(\epsilon_N X_1^N)$ would vanish symmetrically since   
pairs of $X_1$ would necessary have the same $SU(P)$ index. 
Thus, for $N =  m P$,  $m$ different 
objects  up to $ X_{m}  = Z^\dagger (Y  Y^\dagger)^{m-1} Y $ are needed.}
However, unless $N=2 P$ this cannot be 
contracted  into a  $P$-singlet.
We thus need to consider the least common multiplier 
$L\equiv \mathrm{lcm}(P,N)$,  in terms of which the structure of these operators is 
\begin{equation}
    \label{eq:O2Xn}
\mathcal{O}_{I\!I}(X_n) \sim (\epsilon_N)^\frac{L}{N} (\epsilon_P)^\frac{L}{P} 
\!\left(X_1^P \dots X_\mathcal{F}^P\,   
X_{\mathcal{F}+1}^{N-\mathcal{F} P}\right)^\frac{L}{N}\!\!\!\!\!
\end{equation}
where $\mathcal{F}\equiv \mathrm{floor}(N/P)$ denotes the greatest integer 
less or equal to $N/P$.  
Operators of this type  preserve the symmetry defined by $\mX(X_n)=0$,
that is $U(1)_{\xi}$ of \eqn{eq:U1a}, while they break $U(1)_{\xi'} $. 
However, the dimension  $D(L)= (L/N) (\mathcal{F}+1)(2 N - \mathcal{F} P)$  
 grows rapidly with $L$  (for $N=4$ and $P=3$,  $D=30$ !) so that in most   
cases $U(1)_{\xi'} $ breaking  remains an academic issue. 
The dimension of the effective operators of lowest order 
that break respectively  $U(1)_{\xi}$ and $U(1)_{\xi'}$
are given  in Table \ref{tab:Dim}.

\Sec{Vacuum structure of the operators} 
\label{sec:minima}
The PQ solution 
is endangered when the  minimum of the axion potential  is shifted away 
from the one  selected by the non-perturbative  QCD effects.   
Therefore, operators that break explicitly $\UPQ$  in the Lagrangian but  
have vanishing VEVs are harmless, since they do not contribute to determine the minimum.
Thus we need to  study the behaviour of 
$\vev{\mathcal{O}}$, $\vev{\mathcal{O}'}$ at 
the potential minimum.  Let us consider the renormalizable potential for  $Y$. It reads
\beq
\label{eq:V0Y}
V(Y)= \kappa \left(T-  \mu^2_Y 
\right)^2 + \lambda \,A \,, 
\eeq
where $T$ and $A$ are the two 
invariants introduced above, we require  
$\kappa>0$ and  $\lambda > -\frac{2N}{N-1}\kappa$ 
to ensure a potential bounded from below,  
and  $\mu_Y^2>0$ to trigger SSB. 
Let us  write  $Y(x)$
 in its singular value decomposition (SVD):
\begin{equation}
\label{eq:Yc}
\frac{\sqrt{2}}{v_Y}\,Y  = \mathcal{U} \,\hat Y \, \mathcal{V}^\dagger
= U  \hat Y  e^{i \varphi_Y}   V^\dagger \quad \longrightarrow 
\quad \hat Y e^{i \varphi_Y}\,,  
\end{equation}
where $v_Y=\sqrt{2 \vev{T}}$,  
$\mathcal{U}$ and   $ \mathcal{V}$  are  $U(M)$ and $U(N)$ unitary matrices, $U$ and $V$ are  
the corresponding special unitary 
 ($\det (U, V)\!=\!+1$), 
$\varphi_Y \equiv a_Y(x)/v_Y = \frac{1}{M} \arg\det \mathcal{U}- \frac{1}{N}\arg\det\mathcal{V} $
is the NGB  of the global  $U(1)_{\xi_Y}$, and 
 $\hat Y$ is the  matrix of real  non-negative singular values, which 
can be taken to lie in the diagonal upper $N\times N$ block,  while all   
 other entries 
 vanish. 
We will  henceforth  denote as $Y|_{N\uparrow}$   the $N\times N$ 
 {\it upper} left block of a matrix $Y$.
The last form  in \eq{eq:Yc} is obtained 
by gauging away $U(x)$ and $V(x)$. In this gauge 
the two invariants read:
\begin{equation}
    \label{eq:TA}
    T(\hat Y) =  
    \sum_{i=1}^N y_i^2, \qquad A(\hat Y) 
    =      \sum_{i<j} y_i^2 y_j^2\,.  
\end{equation}
It is now easy to identify the vacuum configurations $\hat Y^c\equiv \vev{\hat Y}$ that 
minimize $V(Y)$~\cite{Nardi:2011st}:
$T$ is blind to specific orientations of $\hat Y$ in field space.   
This is because it carries a  $SO(2\!\times \!M\!\times\! N)$ symmetry much
 larger than $\G{MN}$  that allows to rotate different configurations  into  each other. 
Adopting the classification of ref.~\cite{Fong:2013dnk}  it is
a `flavour irrelevant' operator. 
The structure of $\hat Y^c$ is then determined by  the extrema of $A (Y)$.
Since $A$ is non-negative,  its minimum occurs at $\langle
A\rangle=0$, that is when all $y_i$'s but one vanish. The maximum 
instead occurs at the point of enhanced symmetry $y_i^c = 1/\sqrt{N}, \ \forall i$. 
The sign of $\lambda$ thus determines which minimum is selected. 
We take $\lambda <0$ 
so that  $Y^c|_{N\uparrow} = \diag(1,1,
\dots,1)/\sqrt{N}$.   The little group is 
 $\mathcal{H}= SU(N)_V \times SU(M-N)$ with 
 $ SU(N)_{V}$  the `diagonal' combination of 
$SU(N)$ and of  $SU(N)' \subset SU(M)$,  
while   the value of $\varphi^c_Y$ 
is left undetermined.
As regards the renormalizable potential for  $Z$,  in case (I)  it has  the 
form \eqn{eq:V0Y} (with $\mu_Y\to \mu_Z$) but with $A(Z_\alpha)=0$.  
In the SVD \eqn{eq:Yc}  
 $V\to V_Z = I$ while $\hat Z$ has a single non-zero entry in some row $\alpha$ 
with VEV $z_\alpha^c=1$. 
In case (II)  $V_Z$ can be gauged away via a   $SU(P)$ transformation, so that 
$\frac{\sqrt{2}}{v_Z} Z \to  U_Z  \hat Z e^{i\varphi_Z}$ where $\hat Z$ has $P$ singular values located in 
different rows/columns.  For $\lambda_Z<0$ the potential is lowered when  
$\vev{A(Z)}$  is maximum, which corresponds 
to    $z_a^c = 1/\sqrt{P},\, \forall a$.
The relative orientation of  $\vev{Y}$ and $\vev{Z}$ is determined by  
the $D=4$  Hermitian operator
\begin{equation}
\label{eq:Hermit}
\mathcal{O}_{ZY}  = \Tr(Z^\dagger Y Y^\dagger Z)\,.  
\end{equation}
If the coupling is negative,    
the potential is lowered  when $\vev{ \mathcal{O}_{ZY}} $ is  maximum. 
Since $\vev{Y Y^\dagger}|_{N\uparrow} \propto I_{N\times N}$ with all 
other entries vanishing (in particular in the lower $P\times P$ block) 
this occurs  when  the $P$ entries  $z_a$  fall in the upper $N$ positions of  $\hat Z^c$,
while $U^c_Z $, restricted to the block corresponding to these entries, is unitary.
 Thus $\vev{Y}$ and $\vev{Z}$ get maximally aligned, and 
in this case all  $\epsilon_M$-type of operators $\mathcal{O}'$ vanish on the vacuum.
If the coupling is positive, then  $\vev{\mathcal{O}_{ZY} }  \to 0$ which is obtained when  
the entries $z_a$ fill the lower $P$ positions of  $\hat Z^c$, 
and only $U^c_Z|_{P\downarrow} \subset U^c_Z$ is non-trivial (i.e. with off-diagonal entries). 
The two VEVs 
are maximally misaligned, which implies that all $\delta_M$-type of operators have $\vev{\mathcal{O}} = 0$. 
$U^c_Z|_{P\downarrow}$  is unitary but  otherwise undetermined. However, in case (II)  the  $D=M$ 
operator $\mathcal{O}'_{II}$  \eqn{eq:O2YZ} is always allowed, and    
its VEV  would lower the potential proportionally to $\propto  |\vev{\mathcal{O}'_{II}}|$ 
that  is  maximum for  $U^c_Z|_{P\downarrow} \to  I_{P\times P}$.  \\ [-14pt]
 
\label{sec:SU4SU2}
\Sec{A $\mathbf{SU(4)\times SU(2)}$ model} 
As a concrete application of our  study let us outline a model in which 
the PQ symmetry arises automatically and remains perturbatively exact 
(details of the  phenomenology will be discussed elsewhere).
Case (I) with $M\!-\!N>1$   is particularly favorable, since it does not allow for $\epsilon_M$-operators 
that could endanger the anomalous $U(1)_{\xi}$ (see Table~\ref{tab:Dim}).
The minimal  symmetry of this class is  $\G{42}=SU(4)\times SU(2)$. We take $Y\sim (4,\bar 2)$, $Z\sim (4,1)$,
$\mathcal{Q}_L\sim (4,1)$, $Q_R\sim (1,2)$ and $q_R^a = (1,1)$ (with $a=1,2$). 
The {\it flavor relevant} scalar terms and  the quark Yukawa operators  are
\begin{eqnarray}
\label{eq:SU4SU2}
V_{f} &=&  
- \lambda A(Y)
+ \eta \,\mathcal{O}_{ZY}  + \left[\eta_I \mathcal{O}_I^{(6)} + h.c.\right]  \,, \\
V_q &=&  \kappa_Q \bar \mQ_L Y Q_R + \sum\nolimits_{a=1,2} \kappa_{a} \bar \mQ_L Z q^a_R  +  h.c. 
\end{eqnarray}
with $\lambda,\eta >0$.~$A(Y)$  drives  $\hat Y \!\to\! \hat Y^c|_{2\uparrow} \sim \diag(1,1)$ 
at the minimum,  while  $\mathcal{O}_{ZY} $
misalignes  
$\vev{Z}\sim (0,0,z_1,z_2)^T$ and $\vev{Y}$.   
 $\G{42} \to SU(2)_V$ and all the quarks are massive.
As regards  the global symmetries  $U(1)_{\xi_Y}\times U(1)_{\xi_Z} = U(1)_{\xi}\times U(1)_{\xi'}$,  
the  $D=6$ operator $\mathcal{O}_I^{(6)}$ preserves  $U(1)_\xi$  (see  \eqn{eq:O1Xn})   and breaks   
$U(1)_{\xi'}$. However,  VEVs misalignment implies $\vev{\mathcal{O}_I^{(6)}}=0$ which yields  
two NGB: 
\begin{equation}
\label{eq:NGB}
a  = \frac{1}{v_a}  \left(v_Y a_Y + v_Z  a_Z\right), \ \  
a'= \frac{1}{v_a}  \left(v_Y a_Y - v_Z a _Z\right), 
\end{equation} 
where $v^2_a = v^2_Y+v^2_Z$ and,
given that all the fields have the same periodicity,  
we have set $\mX_Y=\mX_Z=1$.
 $a(x)$ gets a mass $m_a \sim m_\pi f_\pi / f_a $ from the QCD anomaly,
 with $f_a = v_a/|2\mathcal{N}|$ and  $|2\mathcal{N}| =  2(\mX_Y + \mX_Z ) =4$. There 
 are, however, only two domain walls because under the  $\mathbb{Z}_2$ center of $SU(2)_V$ 
 $\vev{a} \to \vev{a}+\pi$. At this stage 
 $a'(x)$ remains massless. However,  considering that breaking $U(1)_{\xi'}$ does not  
imply breaking the gauge symmetry,  it might acquire a mass 
{\` a la} Coleman-Weinberg \cite{Coleman:1973jx} once all the effects, including those  
of the fermions, are included in the effective potential. \\ [-14pt]

\label{sec:clockwork}
\Sec{A gauge symmetry for a clockwork  axion}
We now discuss a  
construction based on rectangular gauge symmetries that enforces 
a mechanism for a highly protected  `clockwork'  $\widetilde{U}(1)_{\rm PQ}$.  
Although we use suggestive names for some group factors, this  should 
 be regarded as a toy model  not  intended to describe real phenomenology. 
 
Consider the gauge group $U(1)_\mathcal{Y} \times [SU(2)\times SU(3)]^{n+1}$. We  call  
 $U(1)_\mathcal{Y}$   {\it hyperchage},  and  the first  
 $SU(2)\times SU(3)$  {\it isospin} and {\it flavor}. 
We introduce  three sets of quarks  in the fundamental of color    
transforming  under these factors as $Q_L \sim (2,3)_{\frac{1}{3}}$,
$u^a_R \sim (1,1)_{\frac{4}{3}}$, $d^a_R \sim (1,1)_{-\frac{2}{3}}$  ($a=1,2,3$)
(we leave understood that gauge anomalies 
are compensated by suitable  sets of `leptons')  
and   two scalar multiplets $Y_{d,u}\sim (3,\bar 2)_{\pm1}$ 
which acquire VEVs $\vev{T(Y_{d,u})} =v_{d,u}^2/2$. 
The Yukawa Lagrangian reads:
\begin{eqnarray}
\label{eq:LY}
\mathcal{L}_q = - \sum\nolimits_{a=1}^3 \left(\kappa_u^a \bar Q_L Y_u u^a_R + \kappa_d^a \bar Q_L Y_d d^a_R\right) + h.c,\qquad 
\end{eqnarray}
where  $\kappa^a_{u,d}$ are  coupling constants. 
Note that a coupling $(\epsilon_2 Y_u Y_d)_{\alpha\beta}$  is forbidden  because of  unsaturated 
flavor indices, so that the potential involving  the two scalars has  
the form 
$V(Y_u^\dagger Y_u,Y_d^\dagger Y_d)$ and carry   
an accidental global symmetry $U(1)_{\xi_u}\times U(1)_{\xi_d} = U(1)_{\mathcal{Y}}\times U(1)_\xi$.   
Orthogonality with hypercharge $\mathcal{Y}_u  \mX_u  v^2_u +   \mathcal{Y}_d \mX_d v^2_d = 0 $ 
fixes  the ratio of the  $U(1)_\xi$ charges of the scalars as  $ \mX_u/\mX_d = v^2_d/v^2_u$,   
and we normalize their sum to $\mX_u +\mX_d 
=2$. We now add two sets of  
hyperchargeless fields 
$\Sigma_p, Y_p$ ($p=1,\dots,n$) which  transform under the additional gauge factors. For  $SU(3)\times SU(2)_1 \times SU(3)_1$ 
we add ${\Sigma_{1}}_{\alpha}^{\alpha_1 i_1} \sim (3,\bar {2}_1,\bar {3}_1)$ and ${Y_{1}}_{\alpha_1 i_1}   \sim ( 1, 2_1, 3_1)$, 
and for the successive factors  $\Sigma_{p}  \sim (3_{p-\!1},\bar {2}_p,\bar {3}_p)$ and   $Y_p   \sim ( 1, 2_p, 3_p)$ with $p>1$.
This allows to write a chain of  $n$  renormalizable operators 
\begin{equation}
\label{eq:string}
\hspace{-0.10cm}
(\epsilon_3 \epsilon_2 Y_u Y_d \Sigma_1)^{\alpha_1 i_1} {Y_{1}}_{\alpha_1 i_1} \!+\!
\sum_{p=2}^n  (\epsilon_3 \epsilon_2 Y_{p-1}^2 \Sigma_p)^{\alpha_p i_p} {Y_{p}}_{\alpha_p i_p}\!\!.\!\!\!
\end{equation}
For each field $\Sigma_p$ there is an operator 
 $ \epsilon_2^3 (\epsilon_3 \epsilon_{3'} \Sigma_p^3) (\epsilon_3 \epsilon_{3'} \Sigma_p^3) $
of dimension $D=6$  which, 
together with the operators in \eqn{eq:string}, breaks the  global symmetry 
$U(1)_\xi \times \left[U(1)_\Sigma \times U(1)_Y\right]^n$ to   $ \widetilde{U}(1)_{\rm PQ}$,   
under which   $\tilde \mX_{Y_p} =(- 2)^p$ and $\tilde \mX_\Sigma = 0$. 
Let us now assume that all  dimensional parameters in the scalar potential have values of  order   
$v_{u,d}$ so that there is no large scale in the model.   
The  operators in \eqn{eq:string}  have  multifold effects. First, non vanishing VEVs 
would lower the potential by an amount $\sim |\vev{YY\Sigma' Y'}|$ so that the  VEVs of the fields tend 
to align in specific directions.  The combination 
$\epsilon_2 Y_u Y_d$ 
in the first operator \eqn{eq:string}
misaligns $\vev{Y_{u}} $ and $\vev{Y_d}$ in isospin space, in such a way 
 that after $U(1)_{\mathcal{Y}}\times SU(2)$ breaking a $U(1)$ gauge factor is preserved in the usual way. 
 At the same time   $\epsilon_3 Y_uY_d \Sigma_1$ rotates    
 $\vev{\Sigma_1}$ in the direction  in flavor space  orthogonal to the  plane $\vev{Y_u}$-$\vev{Y_d}$ while, 
because of   $\delta_{2_1}$ and $\delta_{3_1} $ index-contraction,  
$\vev{\Sigma_1}$ and $\vev{Y_1}$ tend to get aligned in $SU(3)_1\times SU(2)_1$ space. 
Isospin breaking provides  a negative mixed-term $ -v_u v_d \Sigma_1 Y_1$ and thus 
there are regions in parameter space  where these two fields acquire a  VEV proportional to $v_{u,d}$ even if 
their squared masses  are non-negative. Hence, regions exist in which all the VEVs of the chain vanish if 
isospin is unbroken and $v_{u,d}\to 0$.
Let us verify if  $ \widetilde{U}(1)_{\rm PQ}$ remains preserved by  higher order operators. 
For each pair $(\Sigma_{p+1},Y_{p+1})$ let us define 
$(X_n)_{\gamma_p}=[\Sigma (\Sigma^\dagger \Sigma)^{n-1} Y]_{\gamma_p}$
with $n=1,2,\dots$.   It is indeed possible to write $ \widetilde{U}(1)_{\rm PQ}$ breaking operators like 
$\epsilon_{3_p} X_1 X_2 X_3$ etc. 
However,  since  $\vev{\Sigma_{p+1}}$ is orthogonal 
in  $SU(3)_p$ space to the plane $\vev{Y_p}_\alpha$-$\vev{Y_p}_\beta$, it has only  
one non-zero $\gamma_p$ component, and thus  all these  
operators vanish on the vacuum. 
Thus the accidental  $ \widetilde{U}(1)_{\rm PQ}$  
is perturbatively exact, and  is broken by the QCD anomaly  with $|2\mathcal{N}| =  3(\mX_u + \mX_d ) =6$. 
The corresponding  NGB  is\nobreak
\begin{equation}
\label{eq:axion}
\tilde a(x) = \frac{1}{v_a} (v_u a_u + v_d a_d + \sum_{p=1}^n v_p a_p ),
\end{equation}
where $v_{u,d,p}$  and $a_{u,d,p}$  are the VEVs  and orbital modes and of $Y_{u,d,p}$ 
and $v_a^2 = 
\mX_u^2 v_u^2 + 
\mX_d^2 v_d^2 + 
\sum_p \mX_p^2 v_p^2  \approx \frac{v^2}{3}4^{n+1}$,  where 
the approximation holds if  all  $v_{u,d,p}\approx v$. 
If we now take the VEVs that break  isospin  and PQ symmetries  at $v\sim 100\,$GeV, then for $n\sim 20$ the radius of the 
axion compact  space is boosted to  $v_a \gsim 10^8\,$GeV  without the need of introducing  any large     fundamental 
parameter.

\label{sec:conclusions}
\Sec{Conclusions}
The   `origin' and  `quality'  problems of the PQ symmetry  can be  solved by assigning the scalar multiplets hosting 
the axion to representations of  semi-simple gauge groups with a `rectangular' structure.  No group 
factors  of large degree are required, which renders the solution particularly elegant.
It should have not gone unnoticed that such constructions require  that (exotic)  
quarks must replicate, with some `generations'  obtaining a mass  from different VEVs than 
others.   Admittedly, the embedding into the SM  of  rectangular  symmetries to play the role of  
flavor symmetries appears to be a challenging undertaking, but hopefully not  insurmountable.   
Succeeding in this venture  might uncover unexpected implications for the SM  flavor problem.

\Sec{Acknowledgments}
We acknowledge conversations with G. Grilli di Cortona. 
The authors are supported by the INFN Iniziativa Specifica, 
Theoretical Astroparticle Physics (TAsP-LNF).



%

\end{document}